\documentclass[10pt]{article}
\usepackage[OE]{express}

\usepackage{combelow} 
\usepackage{float}
\usepackage[normalem]{ulem}

\begin{document}

\title{Pulsed single-photon spectrometer by frequency-to-time mapping using chirped fiber Bragg gratings}

\author{Alex O. C. Davis,\authormark{1} Paul M. Saulnier,\authormark{2} Micha{\l} Karpi\'{n}ski,\authormark{1,3,*} and Brian J. Smith\authormark{1,4}}

\address{\authormark{1}Clarendon Laboratory, Department of Physics, University of Oxford, Parks Road, Oxford, OX1 3PU, United Kingdom\\
\authormark{2}Department of Physics, Gustavus Adolphus College, 800 West College Avenue, Saint Peter, Minnesota, 56082, USA\\
\authormark{3}Faculty of Physics, University of Warsaw, Pasteura 5, 02-093 Warszawa, Poland\\
\authormark{4}Department of Physics and Oregon Center for Optical, Molecular, and Quantum Science, University of Oregon, Eugene, Oregon 97403, USA}

\email{\authormark{*}mkarp@fuw.edu.pl} 



\begin{abstract}
A fiber-integrated spectrometer for single-photon pulses outside the telecommunications wavelength range based upon frequency-to-time mapping, implemented by chromatic group delay dispersion (GDD), and precise temporally-resolved single-photon counting, is presented. A chirped fiber Bragg grating provides low-loss GDD, mapping the frequency distribution of an input pulse onto the temporal envelope of the output pulse. Time-resolved detection with fast single-photon-counting modules enables monitoring of a wavelength range from $825$~nm to $835$~nm with nearly uniform efficiency at $55$~pm resolution ($24$~GHz at $830$~nm). To demonstrate the versatility of this technique, spectral interference of heralded single photons and the joint spectral intensity distribution of a photon-pair source are measured. This approach to single-photon-level spectral measurements provides a route to realize applications of time-frequency quantum optics at visible and near-infrared wavelengths, where multiple spectral channels must be simultaneously monitored.
\end{abstract}



\section{Introduction}

Optical experiments involve three essential steps, namely generation, manipulation and detection of light. In the case of quantum optics, non-classical states of light enable performance beyond that possible with classical resources alone in a range of applications such as precision metrology \cite{LIGO, demkowicz:12}, and quantum key distribution \cite{zhong:15,nunn:13}. Here, information can be encoded in a variety of degrees of freedom, such as the amplitude and phase of a single field mode \cite{braunstein:05}
, the polarization \cite{pittman:02}{
, or the transverse-spatial mode \cite{sasada:03} of a single photon. Pulsed wave-packet modes of light, in which the arrival time and central frequency of a pulse can be used as robust information carriers, are ubiquitous in classical telecommunications due to their natural compatibility with integrated-optical platforms and unprecedented information capacity. Pulsed modes have recently gained significant interest in quantum optics both for quantum information processing applications \cite{humphreys:14, brecht:15} and metrology \cite{lamine:08}. To harness the advances enabled by single-photon spectral mode encoding requires the ability to directly measure the frequency of an ensemble of identically prepared individual photons.

Here, a method to simultaneously monitor multiple spectral channels of a pulsed single-photon source is proposed and demonstrated. Our approach utilizes the dispersive Fourier transform concept \cite{tong:97,torres:11,goda:13} to map the frequency of a temporally-short optical pulse onto the temporal envelope of the output pulse. We use advances in chirped fiber Bragg grating (CFBG) technology \cite{canning:08, ibsen:00} to achieve large linear group-delay dispersion (GDD) across a broad frequency range outside the telecommunications spectral region. This technique allows rapid data acquisition rates of high-resolution spectral measurements. The concurrent monitoring of many frequency modes enables multiple bits of information to be encoded per single photon. To demonstrate the wide-ranging applicability of this method, we present spectral measurements of a heralded single-photon occupying a double pulse mode that displays spectral interference fringes with unprecedented resolution outside the telecommunications wavelength range, and the joint spectral intensity (JSI) from a degenerate two-photon source. Our device is, to the best of our knowledge, the first spectrometer to use a CFBG in conjunction with fast single-photon-level detection and coincidence measurement to achieve spectrally-resolved single photon detection with resolution unprecedented away from telecommunications wavelengths. The technique shows comparable performance with that achieved at telecommunications wavelengths using fiber-assisted single-photon spectroscopy \cite{kuo:16}.

\begin{figure}[t!]
\centering
\includegraphics{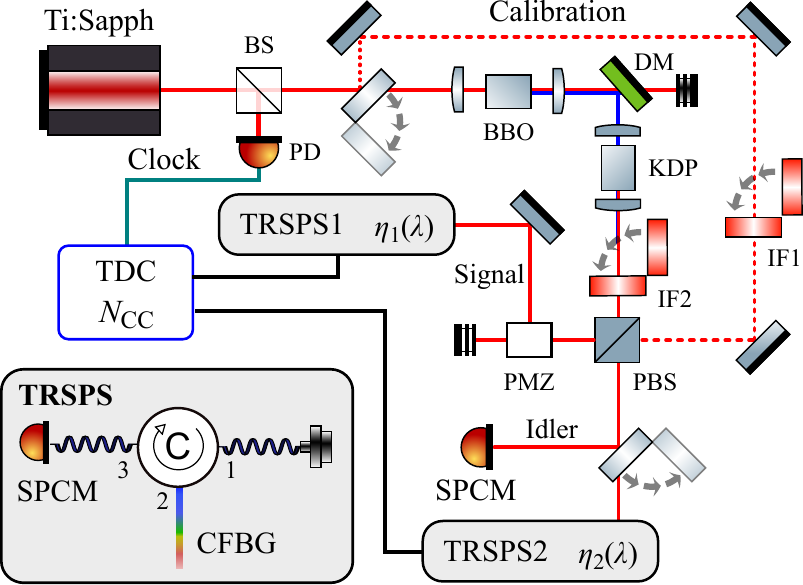}
\caption{Experimental setup. Ti:Sapph, Ti:sapphire femtosecond oscillator; BBO (KDP), beta barium borate (potassium dihydrogen phosphate) crystal; (P)BS, (polarizing) beam splitter; PD, fast photodiode; TDC, time-to-digital converter; DM, dichroic mirror; IF1(2), bandpass filters; PMZ, polarization Mach-Zehnder interferometer; TRSPS1(2), time-resolved single photon spectrometer; SPCM, single photon counting module. For calibration the TRSPSs were fitted with linear PDs. See text for details.}
\label{fig:setup}
\end{figure}

Two general approaches can be taken to acquire the spectrum of a beam of light. One approach involves scanning an optical element that maps spectral information onto the amplitude of the beam followed by direct intensity detection. This forms the basis of spectral measurements using Fabry-P\'erot etalons and Fourier-transform spectrometers. The alternative approach couples spectral modes with another set of modes such as transverse spatial modes. This technique underlies the standard spectrograph in which an angular dispersion element, such as a prism or grating, maps the spectral modes of an input optical beam onto a range of transverse spatial frequencies that can be transformed into localized spots by a lens and monitored with an array of intensity detectors. Each approach has different advantages, for example methods that involve scanning require only a single detector and achieve high spectral resolution. However, scanning methods suffer from the time needed to acquire full spectral information and can be limited in the overall spectral range that can be readily accessed. The spectra of heralded single-photon and two-photon states have been realized using scanning monochromators \cite{kim:05} and Michelson interferometers \cite{wasilewski:06}. A significant drawback to scanning measurements is that this approach does not allow access to all spectral modes simultaneously, reducing the information content available in a single measurement outcome. This fact renders scanning measurements incompatible with real-time information processing and communications.

In principle, mapping frequency modes of a beam onto the transverse spatial modes allows simultaneous monitoring of all spectral channels with a spatial array of detectors. However, this approach is difficult to realize with heralded single photons and two-photon states where coincidence events must be registered, owing to challenges in fabrication of large, low-noise and efficient arrays of independent single-photon-counting detectors \cite{zappa:07}. The frequency distribution of pulsed light can be mapped onto temporal modes by a medium with high GDD and monitored with a single intensity detector with high temporal resolution \cite{tong:97,goda:13}. At the single-photon level this method was first used to measure the JSI of a biphoton source at telecommunications wavelengths \cite{avenhaus:10} where a $3.3$~km long dispersion compensating fiber introduced the necessary GDD. Subsequent works \cite{kuo:16} have improved the resolution by using more dispersive fibers. Within the visible spectral domain significant losses in optical fibers make this approach infeasible. This challenge can be overcome by using low-loss chirped fiber Bragg gratings (CFBGs) as the dispersive medium, as demonstrated for classical light pulses \cite{muriel:99, azana:00}. In such a scheme, the GDD and wavelength range are mutually constrained by the maximum optical delay introduced between the two extremes of the CFBG reflection window (determined by its length and refractive index). For a given length of grating, a greater GDD (and hence greater spectral resolution) results in a narrower detection bandwidth. This trade-off can be eased by using longer CFBGs or by splicing multiple CFBGs with adjacent reflection windows in succession. For all time-of-flight spectrometry schemes, extending the maximum optical delay of the dispersive element in this manner is ultimately limited by the repetition rate of the experiment, since one must be able to discern from the time of arrival exactly which cycle of the experiment clock the pulse originated. Here we combine time-of-flight spectrometry with CFBGs and time-resolved heralded single-photon detection, presenting a high-resolution single-photon spectrometer that allows simultaneous monitoring of multiple spectral modes using a single detector at high repetition rates and outside the telecommunications wavelength range. 

\section{Experiment}

The time-resolved single-photon spectrometer (TRSPS) is constructed from a fiber-based optical circulator (FCIR-2285-14B181, Haphit) with specified insertion loss of 1.7 dB spliced to a chirped fiber Bragg grating (CFBG, TeraXion) operating in reflection. The CFBG reflectivity $R \approx 0.5 \pm 0.05$ is nearly constant across a wavelength range of $825-835$~nm. The circulator output is directed to a fast SPCM as depicted in Fig.~\ref{fig:setup} (inset). The pulse to be characterized is launched into port $1$ of the circulator and directed to port $2$, where it is reflected by the CFBG, which provides nearly constant GDD. The reflected pulse emerges from port $3$ of the circulator and is detected by a fast SPCM (PDM, Micro Photon Devices) with approximately $\Delta t = 50$~ps at full width half maximum (FWHM) timing jitter and specified detection efficiency at $830$~nm of approximately 10\%. The SPCM signal is monitored by a time-to-digital converter (TDC, PicoQuant, HydraHarp 400) triggered by a fast photodiode signal that samples the laser pulse train acting as the experiment clock. The TDC has a timing resolution of $<12$~ps RMS, below the detector timing resolution. The trigger photodiode (Alphalas, UPD-15-IR2-FC) has approximately $15$~ps rise time and timing resolution less than $1$~ps. The TDC records a list of times between trigger events and SPCM signals, $\{\tau_j\}$, which are then converted into a histogram of coincidence counts as a function of these times, $N_{\rm{CC}}(\tau)$. The histogram time bin size of $32$~ps is chosen to be smaller than the SPCM timing jitter. 

The wavelength-dependent time of arrival relative to a pulse centered on the central wavelength of the CFBG ($\lambda_0 = 830$ nm) is $t = D(\lambda-\lambda_0)$, where $D$ is the GDD of the CFBG. We assume that the original temporal support of the pulse is considerably less than the SPCM timing jitter of 52 ps such that the pulse undergoes complete frequency-to-time mapping on the timescale of the detector resolution. Since the transform-limited duration of our pulses is  less than 1 ps this is a valid approximation in our demonstration. Thus the spectral probability distribution of heralded single photons, $S(\lambda)$, from this histogram will be 
\begin{equation}
S(\lambda) = N_{\rm{CC}}(D \lambda - \delta \tau) / \eta (\lambda),
\label{eq1}
\end{equation}
where $\eta(\lambda)$ is the wavelength-dependent efficiency of the TRSPS, including contributions from both the circulator-CFBG setup and the efficiency of the SPCM. The time offset $\delta \tau$ depends upon optical and electronic delays in the experiment and sets the central wavelength of the device response. Thus, to determine the spectrum of pulsed single photons, the time $\delta \tau$, GDD $D$, and efficiency $\eta (\lambda)$ must be ascertained.

The temporal point-spread function of the detector-TDC setup has FWHM width $\Delta t = 52$~ps, which is determined by sending single photons directly into the setup, bypassing the CFBG, and measuring the resulting temporal distribution. The CFBG introduces approximately $950~\mathrm{ps}/\mathrm{nm}$ of GDD to the optical pulse train centered about $830$~nm wavelength, resulting in a frequency-dependent group delay that allows direct measurement of the pulse spectrum with time-resolved detection. The spectral resolution is set by the ratio of detector timing resolution and GDD, $\Delta \lambda = \Delta t / D = 0.055$~nm FWHM. The total optical transmission through the device was measured by launching broadband classical pulses into the spliced circulator-CFBG combination and monitoring the input and output spectral intensity with a standard grating spectrometer (Andor Shamrock 303i) with $0.054$~nm resolution. The total loss is approximately $10$~dB across the $825-835$~nm wavelength range, comprising $3.4$~dB from the two passes through the circulator, $3$~dB from the CFBG and the remainder from mode-mismatching losses in the fiber splices. These losses are technical in nature and could be readily reduced with further development. By comparison, if single-mode fiber is used instead of a CFBG as the dispersive element, for a scheme achieving comparable resolution one would expect the losses from attenuation alone to be approximately $30$--$35$~dB, using typical values of the loss and GDD of single-mode fiber at $830$~nm, $3.5$~dB/km and $0.1$~ns/(nm$\cdot$km), respectively. If direct filtering were used to sample the spectrum, as in a standard scanning monochromator, one would have to scan up to $200$ spectral bins to achieve comparable resolution across the whole $825-835$~nm range. Assuming this could be done without loss and using an SPCM of the same detection efficiency, this would increase single-photon spectrum measurement times by up to a factor of $20$. Although we note that without the need for precise timing resolution, more efficient SPCMs could be used to improve this. Such an approach would also not allow the concurrent monitoring of multiple spectral modes and hence is of greatly reduced use for numerous applications, such as communications.

\begin{figure}[t!]
\centering
\includegraphics[width=\columnwidth]{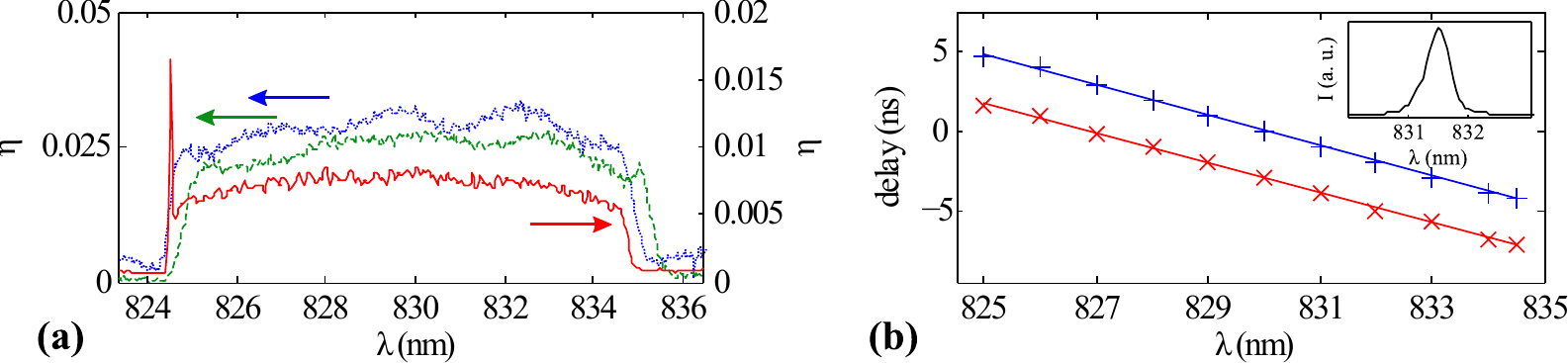}
\caption{(a) Spectral response functions of the three TRSPS setups. Response is close to zero outside of the $825-835$~nm reflection window of the CFBGs. CFBG1 (dotted, blue) and CFBG2 (dashed, green) with PerkinElmer (slow) SPCMs. CFBG1 with MPD (fast, less efficient) SPCM (solid red, right scale). The peak at $824.6$~nm is caused by back-reflection from a fiber splice. This feature is present in the dotted blue dataset as well but is not resolved by the slower SPCM. (b) Wavelength-dependent delays introduced by the CFBGs, obtained using bright laser pulses for TRSPS1 (blue, $+$) and 2 (red, $\times$) with linear fits (solid lines). INSET: Example of narrowband filtered spectrum used for calibration.}
\label{fig:transmission}
\end{figure}

To determine $D$, laser pulses with narrow bandwidth derived from a spectrally filtered Ti:Sapphire oscillator were routed into the device. The pulse train was passed through a $1$-nm-bandwidth interference filter (Andover; IF2 in Fig.\ \ref{fig:setup}), which was tilted such that the narrowband transmitted light could be tuned across the range $825-835$~nm. The spectrum of light emerging from the filter was measured with the grating spectrometer. The output of the circulator was coupled into a fast photodiode (DET$10$C, Thorlabs) and the delay of the signal relative to the experiment clock was measured on an oscilloscope (Tektronix MSO 5104). The measured delay as a function of wavelength is presented in Fig.~\ref{fig:transmission}(b) for two different circulator-CFBG setups. A polynomial fit applied to the data yields a GDD of $938~\mathrm{ps}/\mathrm{nm}$ and $958~\mathrm{ps}/\mathrm{nm}$ in TRSPS 1 and 2 respectively. Higher-order wavelength dependence of the time delay was negligible and is henceforth disregarded. The value of $\delta \tau$ depends on electronic and optical delays and must be determined independently for each experimental configuration. For the calibration setup, shown by the dotted line in Fig.~\ref{fig:setup}, the bright pulses were spectrally filtered at $830$~nm central wavelength and the CFBG-circulator setup attached to a fast SPCM and TDC to complete the TRSPS. The peak in the temporal histogram can then be associated with the central wavelength of the calibration pulse, and with $D$ known, all the single-photon time tags {$\tau_i$} can be attributed to specific wavelength values.  In the subsequent demonstrations involving single photons, where the beam follows a different path (the downconversion path shown by the solid line in Fig.~\ref{fig:setup}), the timing offset $\delta \tau$ is recovered in the same fashion by using narrowband filtration of the single-photon light source itself with filter IF2 and comparing with the spectrum transmitted by the filter, obtained with a long integration time measurement on the standard spectrometer. 

To determine the efficiency $\eta(\lambda)$, the same beam path -- the dotted line in Fig.~\ref{fig:setup} -- is used without the filter IF1. Removing this filter subtracts only $6$~ps of delay, less than the temporal resolution of the SPCM, and therefore to within the resolution of the device the same value for $\delta \tau$ is retained. The spectrum, $I_{\rm{in}}(\lambda)$, of an attenuated classical pulse is measured with the grating spectrometer. This pulse is then directed through the TRSPS setup and the total count rate at the output as a function of wavelength, $N_S(\lambda)$, is calculated from the resulting temporal histogram. This enables characterization of the wavelength-dependent response $\eta(\lambda) = N_S(\lambda)/AI_{\rm{in}}(\lambda)$, where $A$ is a normalization chosen such that $\int \eta(\lambda) d\lambda = H$,
where $H$ is the total heralding efficiency of the setup. Therefore $\eta(\lambda)$ is normalized such that it corresponds to the probability density of detecting a photon at $\lambda$, conditional on a heralding event. The resulting response function $\eta(\lambda)$ for the three TRSPS devices used is shown in Fig.~\ref{fig:transmission}(a).

To demonstrate the versatility of this device, we first measure the spectrum of heralded single photons occupying a double pulse displaying spectral interference fringes. Heralded single photons are generated by collinear, type-II spontaneous parametric down conversion (SPDC) in an $8$-mm-long potassium dihydrogen phosphate (KDP) crystal \cite{mosley:08}. The KDP crystal is pumped by a frequency-doubled (in $0.7$-mm-thick beta barium borate, BBO) Ti:Sapphire laser oscillator operating at $80$~MHz repetition rate with $830$~nm central wavelength and $8$~nm FWHM bandwidth. The orthogonally polarized degenerate signal and idler fields with, respectively, $2$~nm and $8$~nm FWHM bandwidths are separated at a polarizing beam splitter (PBS), with a photodetection event in the idler mode heralding the creation of a photon in the signal mode. To produce a single photon in a double pulse, the heralded photon is sent through an unbalanced polarization Mach-Zehnder interferometer (PMZ) consisting of a half wave plate (HWP) rotating the polarization to $45^{\circ}$ followed by a Soleil-Babinet compensator (SBC) that introduces a controllable group delay, $T$, between horizontal (H) and vertical (V) polarization modes. The delay $T$ was determined from the fringe spacing in the spectral interference pattern of classical light. An additional HWP rotates the H (V) polarization to (anti-)diagonal and a polarizer selects H-polarized components of the field to be transmitted, probabilistically yielding a single-photon double pulse. The double pulse train output was sent to the TRSPS to analyze the spectrum, which was obtained by using the heralding detection event to trigger the TDC and the detection time of the signal photon. The resulting spectral interference pattern, shown in Fig.~\ref{fig:singlephoton}(a), displays an oscillation period of $0.2$~nm, matching the time delay $T = 11$~ps introduced by the SBC. This spectrum, acquired over $300$~s, contains $3\,000$ heralded single-photon detection events.

\begin{figure}[t!]
\centering
\includegraphics{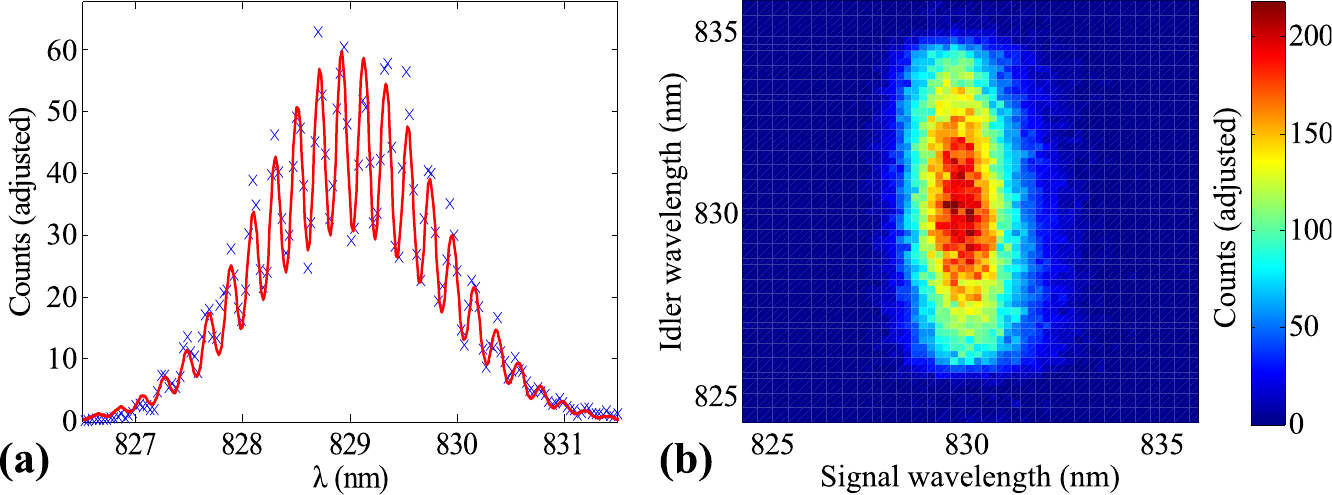}
\caption{(a) Experimental heralded single-photon spectrum in a double-pulse mode. The number of counts is adjusted to account for nonuniform efficiency of the TRSPS whilst conserving the total number of counts. The red curve shows the expected spectrum given the underlying single-photon spectrum and the time delay, with visibility and phase chosen freely in order to give the best fit. To obtain the red curve, a Gaussian fit to the underlying spectrum was obtained from a separate measurement of the photon spectrum; this was modulated by a sinusoidal signal with known period given by the time delay. The visibility of 24\% is primarily reduced by phase instability in the interferometer over the 300 s acquisition time. (b) Measured JSI of SPDC photon pair source. Coincidence counts per bin, adjusted to account for the spectral variation of TRSPS efficiency whilst conserving total count number.}
\label{fig:singlephoton}
\end{figure}

To further demonstrate the utility of the TRSPS, the JSI of the SPDC source was measured. This was achieved by directing the idler photon to a second TRSPS and observing coincidence events between the signal and idler detection events. This spectrum, acquired over $600$~s, contains $28\,000$ joint two-photon detection events. Characterization of the JSI of nonlinear optical sources has recently been demonstrated in the stimulated-emission regime using a bright continuous wave seed in the idler mode to create a bright signal beam that can be resolved conventionally, enabling high spectral resolution \cite{lorenz:14}. However, this approach cannot be applied for more general photon-pair sources. Directly resolving individual photon pairs has been achieved away from telecommunications wavelengths using variants on scanning methods in which most photons are discarded \cite{mosley:08,kim:05}, or the spectral bins are not monitored directly  \cite{wasilewski:06}. The approach described here allows simultaneous monitoring of many spectral modes and therefore far higher count rates and shorter acquisition times. The recently proposed spectral-spatial-temporal mapping setup \cite{poem:16} is conceptually similar to our method, albeit lacks the all-fiber character and achieves much lower resolution over a limited spectral range.

The joint spectral measurement employed more efficient SPCMs (PerkinElmer SPCM-AQ4C) to enable sufficient coincidence count rates. These detectors have quantum efficiency of approximately $0.4$ at $830$~nm central wavelength, which comes at the cost of reduced timing resolution, $\Delta t=200$~ps, and thus lower spectral resolution $\Delta \lambda = 0.2$~nm. The coincidence timing was implemented using a lower-timing-resolution TDC (quTAU, quTOOLS; $81$~ps timing resolution), capable of a higher number of input channels than the fast TDC, which allowed us to monitor two TRSPS triggered by the clock signal. To extract the JSI for the SPDC source each TRSPS was calibrated in the same manner as for the heralded single-photon interferogram experiment, and then time stamps at each SPCM were used to determine the corresponding wavelengths. The normalized JSI is shown in Fig.~\ref{fig:singlephoton}(b). Note that full spectral range of the idler photon is not accessible because the reflectivity of the CFBG is limited to a $10$~nm range about $830$~nm. This could be improved by exchanging the CFBGs by those with broader spectral range.

\section{Summary}
In summary, we have demonstrated a method to provide simultaneous high-resolution monitoring of the spectrum of pulsed modes at the few-photon level in an all fiber set-up outside the telecommunication wavelength range. The approach utilizes frequency-to-time mapping introduced by large second-order dispersion in a chirped fiber Bragg grating followed by fast time-resolved coincident single-photon detection. The calibration of the device can be achieved using a narrowband classical light source and conventional spectrometer. This approach enables conditional high-resolution spectral measurement of light allowing acquisition rates that are orders of magnitude faster than previously achieved. We anticipate this method will find use in a broad range of applications for low-light sensing from quantum foundations to technology, such as time-frequency quantum key distribution.

\section*{Funding}
European Union's (EU) Horizon 2020 research and innovation programme under Grant Agreement No. 665148. National Science Foundation award No. 1620822. United Kingdom Defense Science and Technology Laboratory (DSTL) (contract No.~DSTLX-100092545). National Science Centre of Poland (NCN) (2014/15/D/ST2/02385).

\section*{Acknowledgments} We thank E.\ Poem and M.\ Nejbauer for insightful discussions and D.\ Gauthier for assistance in sourcing the CFBGs.

\end{document}